\documentclass[prl,aps,twocolumn,showpacs,preprintnumbers,amsmath,amssymb]{revtex4}

\usepackage{graphicx}
\usepackage{dcolumn}

\begin{document}
\title{Giant Magnetoresistance Oscillations Induced
by Microwave Radiation and a Zero-Resistance State
in a 2D Electron System with a Moderate Mobility}

\author{A.~A.~Bykov,
	A.~K.~Bakarov,
	D.~R.~Islamov,
	and A.~I.~Toropov}

\affiliation{Institute of Semiconductor Physics, Siberian Division,
Russian Academy of Sciences, Novosibirsk, 630090 Russia\\
e-mail: bykov@thermo.isp.nsc.ru
}


\begin{abstract}
The effect of a microwave field in the frequency range
from 54 to $140$~$\mathrm{GHz}$ 
on the magnetotransport in a GaAs quantum well with AlAs/GaAs superlattice
barriers and with an electron mobility no higher than $10^6$~$\mathrm{cm^2/Vs}$
is investigated. In the given two-dimensional system under the effect of
microwave radiation, giant resistance oscillations are observed with their
positions in magnetic field being determined by the ratio of the radiation
frequency to the cyclotron frequency. Earlier, such oscillations had only
been observed in GaAs/AlGaAs heterostructures with much higher mobilities.
When the samples under study are irradiated with a $140$-$\mathrm{GHz}$
microwave field, the resistance corresponding to the main oscillation minimum,
which occurs near the cyclotron resonance, appears to be close to zero.
The results of the study suggest that a mobility value lower than
$10^6$~$\mathrm{cm^2/Vs}$ does not prevent the formation of zero-resistance
states in magnetic field in a two-dimensional system under the effect of
microwave radiation.
\end{abstract}


\maketitle

Current interest in studying the transport in twodimensional (2D)
electron systems is related to the recent observation of
resistance oscillations in magnetic field that arise in
high-mobility GaAs/AlGaAs heterostructures under the effect of
microwave radiation \cite{bib1}. It was found that these
oscillations are periodic in the inverse magnetic field
($1/B$) with a period determined by the ratio of the microwave
radiation frequency to the cyclotron frequency. The photoresponse
oscillations in magnetic field in a high-mobility 2D system
(such oscillations were predicted more than 30 years ago \cite{bib2})
fundamentally differed from the behavior of photoresponse
in GaAs/AlGaAs heterostructures with lower mobilities \cite{bib3}.
The effect of microwave radiation on the magnetotransport in
GaAs/AlGaAs heterostructures of moderate quality was found
to manifest itself as a photoresistance peak
caused by the heating of the 2D electron gas under the
magnetoplasma resonance conditions \cite{bib4}. Soon after
the first experimental observation of the microwave
radiation-induced resistance oscillations in magnetic
field in high-mobility GaAs/AlGaAs heterostructures,
it was shown that the minima of these oscillations may
correspond to resistance values close to zero \cite{bib5,bib6,bib7}.

This unexpected experimental result initiated intensive
theoretical studies of the aforementioned phenomenon
\cite{bib7,bib8,bib9,bib10,bib11,bib12,bib13,bib14,bib15,bib16}.
However, despite the multitude of theoretical publications,
the mechanisms responsible for the resistance oscillations
under the effect of a microwave field in 2D systems with
large filling factors remain open to discussion.
The role of the mobility of charge
carriers in the manifestation of microwave-induced
zero-resistance states arising in magnetic field in 2D
systems also remains unclear. It is commonly believed
that the mobility should exceed $3\times 10^6 \mathrm{cm^2/Vs}$
\cite{bib17}. As for the experimental studies of the photoresponse to
microwave radiation in 2D systems in classically strong
magnetic fields, such studies, excluding a few of them
\cite{bib17,bib18,bib19,bib20},
are restricted to high-mobility GaAs/AlGaAs
heterostructures with thick spacers and, hence, low
electron concentrations \cite{bib21,bib22,bib23}.

In this paper, we report on the observation of resistance
oscillations periodic in the inverse magnetic field
that arise under the effect of irradiation with millimetric
waves in GaAs quantum wells with a much lower
mobility and a much higher concentration of 2D electrons,
as compared to those reported earlier in
\cite{bib1,bib5,bib6,bib7}.
We experimentally demonstrate that, despite the relatively
low mobility, in the 2D system under study at the
temperature $T = 1.7$~K, the effect of microwave radiation
at the frequency $F = 140$~$\mathrm{GHz}$ gives rise to a close-to-zero
resistance state in magnetic field.

We studied heterostructures with modulated doping,
which represented GaAs quantum wells with
AlAs/GaAs superlattice barriers \cite{bib18,bib18,bib20}.
The width of a GaAs quantum well was $13$~nm. The structures were
grown by molecular beam epitaxy on GaAs substrates.
After a short-period illumination by a red light-emitting
diode, the concentration and mobility of 2D electrons
in our samples were $n_e=8.5\times 10^{11}$~$\mathrm{cm^{-2}}$ and
$\mu=560\times 10^3$~$\mathrm{cm^{2}/Vs}$, respectively.
The measurements were carried out at temperatures of
$1.7$ and $4.2$~K in magnetic fields $B$ up to $0.6$~T.
We used Hall bars with a width of $50$~$\mu\mathrm{m}$ and a
spacing of $250$~$\mu\mathrm{m}$ between the potential 
contact pads. The microwave radiation was supplied to
the sample via a circular waveguide with an inner diameter
of $6$~mm. The output microwave power of the generators
used in the experiment was $P_\mathrm{out} = (4$--$10)$~$\mathrm{mW}$.
The resistance was measured using an alternating current of
$(1$--$10) \times 10^{-7}$~$\mathrm{A}$ with a frequency of
$(0.3$--$1)$~$\mathrm{kHz}$.

\begin{figure}[t]
 \includegraphics[width=3.1in,clip]{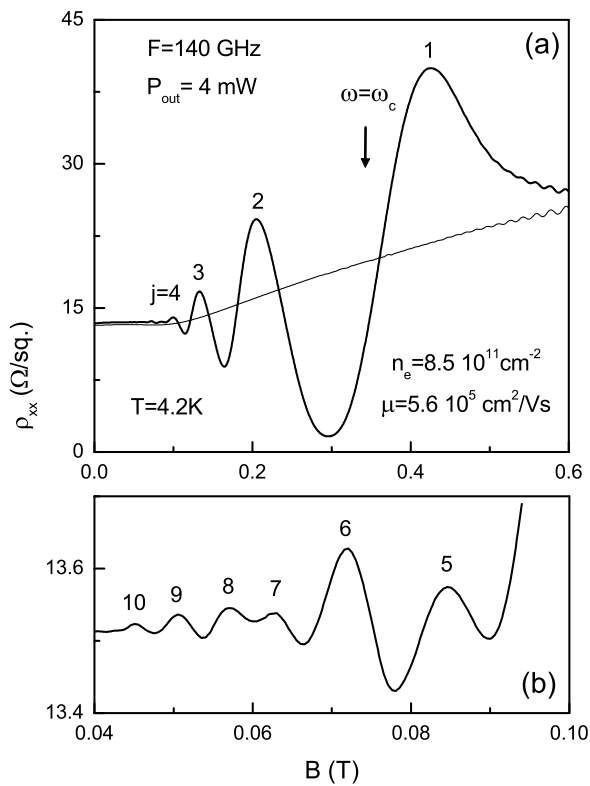}
 \caption{(a) Experimental $\rho_{xx}(B)$ dependences for the 2D
  electron gas in a GaAs quantum well with AlAs/GaAs
  superlattice barriers at $T=4.2$~$\mathrm{K}$ (the thin line) in the
  absence and (the thick line) in the presence of $140$-$\mathrm{GHz}$
  microwave radiation. (b) A detailed $\rho_{xx}(B)$ dependence in
  the presence of microwave radiation in magnetic fields
  below $0.1$~$\mathrm{T}$. The arrow indicates the position of the cyclotron
  resonance. The ordinal numbers of maxima are indicated
  near them beginning from the main maximum.
 }
 \label{fig1}
\end{figure}

Figure~\ref{fig1} shows the $\rho_{xx}(B)$ dependences in the
absence of microwave radiation and in the presence of
radiation at a frequency of $140$~$\mathrm{GHz}$ at a temperature of
$4.2$~$\mathrm{K}$. One can see that the microwave radiation causes
giant oscillations of magnetoresistance. An analysis of
the positions of the first four maxima of these oscillations
shows that they are periodic in inverse magnetic
field. However, the positions of the maxima with numbers
$5$--$10$ deviate from the linear dependence on
inverse magnetic field. A qualitatively similar behavior
was observed earlier for high-mobility GaAs/AlGaAs
heterostructures with a much smaller concentration \cite{bib6}
and was explained by the spin splitting \cite{bib24}. One also
can clearly distinguish specific points where the microwave
radiation does not affect the resistance of the 2D
electron gas. One of such points is indicated by the
arrow in the plot and corresponds to the condition
$\omega=\omega_\mathrm{c}$, where $\omega$ is the microwave radiation
frequency and $\omega_\mathrm{c}$ is the cyclotron frequency. The minimum that is
closest to this point is the deepest of all the minima in the plot.

\begin{figure}[t]
 \includegraphics[width=3.1in,clip]{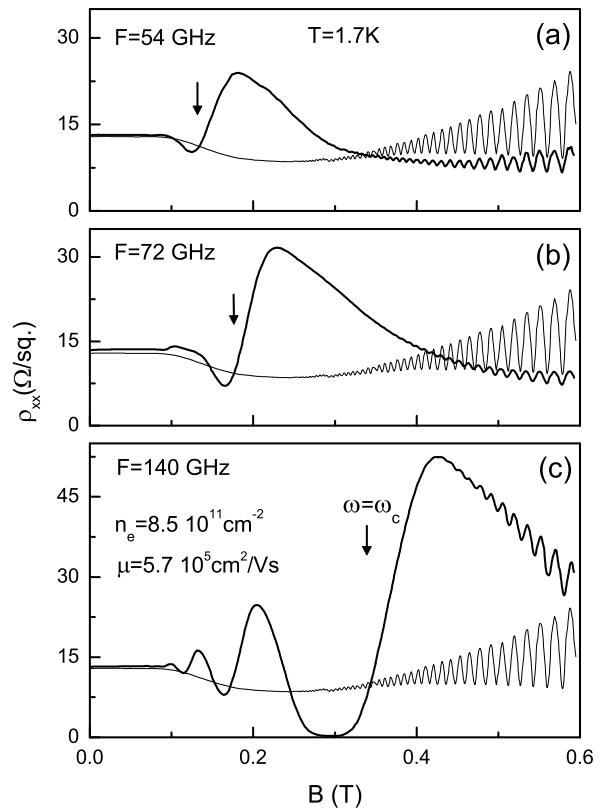}
 \caption{Experimental $\rho_{xx}(B)$ dependences for the 2D
  electron gas in a GaAs quantum well with AlAs/GaAs
  superlattice barriers at $T = 1.7$~$\mathrm{K}$ (the thin line)
  in the absence and (the thick line) in the presence of microwave
  radiation with the frequencies $F = $~(a)~$54$, (b)~$72$,
  and (c)~$140$~$\mathrm{GHz}$. The arrows indicate the positions
  of the cyclotron resonance.
 }
 \label{fig2}
\end{figure}

Figure~\ref{fig2} illustrates the effect of microwave radiation
on the magnetoresistance of 2D electron gas in a GaAs
quantum well with AlAs/GaAs superlattice barriers at a
temperature of $1.7$~$\mathrm{K}$ for three different microwave
frequencies: $54$, $72$, and $140$~$\mathrm{GHz}$. The experimental
dependences presented in this figure show that the point
of intersection of the curves that corresponds to the
cyclotron resonance position is shifted in magnetic field
as the frequency varies. In addition, one can see that, at
the microwave frequency $F = 140$~$\mathrm{GHz}$ and a
temperature of $1.7$~$\mathrm{K}$, the resistance corresponding
to the main minimum located near the aforementioned point takes
a value close to zero.

It should be noted that, when the temperature is lowered from
$4.2$ to $1.7$~$\mathrm{K}$, the $\rho_{xx}(B)$ dependence observed
in the absence of microwave radiation considerably
changes in appearance. On the one hand, the Shubnikov-de Haas
oscillation amplitude increases, and, on
the other hand, the positive magnetoresistance typical
of the given heterostructures at $T = 4.2$~$\mathrm{K}$ changes to
negative magnetoresistance at $T = 1.7$~$\mathrm{K}$.
An analysis of the experimental data shows that the negative
magnetoresistance that appears at the lower temperature is not
described by a quadratic dependence on B: it exhibits a
``shelf'' near $B = 0$ and, hence, cannot be caused by the
electron-electron interaction alone \cite{bib25}. Therefore, we
believe that the negative magnetoresistance observed in
our structures in classically strong magnetic fields
should be qualitatively explained as the result of a
combined effect produced on the transport processes by the
electron-electron interaction and the classical ``memory''
effects \cite{bib26,bib27,bib28}.

For identification of the resistance oscillations
induced by microwave radiation, it is necessary to
determine the positions of the specific points of these
oscillations on the $B$ axis at a fixed frequency $\omega$, i.e.,
the positions of maxima, minima, and points where the
photoresponse is absent. Most of the theoretical
publications allow an exact determination of positions for
only those points where the resistance is unaffected by
microwave radiation. The positions of these points
correspond to the cyclotron resonance and its harmonics
$\omega=n\omega_\mathrm{c}$, where $n$ is a positive integer number.
This theoretical result agrees well with the experimental data
\cite{bib29}, which suggest that the most accurately
determined point of resistance oscillations in magnetic field
is the point corresponding to the cyclotron resonance.
This point lies between the main maximum and the
main minimum, at the intersection of the $\rho_{xx}(B)$
dependences obtained in the absence and in the presence of
microwave radiation.

Such an intersection is observed in our dependences.
From the analysis of our experimental data, it
follows that the position of the intersection is shifted in
magnetic field with varying frequency in accordance
with the displacement of the cyclotron resonance position
calculated by using the effective electron mass in
GaAs ($0.067m_0$). This fact suggests that the nature of
oscillations presented in Figs.~\ref{fig1} and \ref{fig2} and arising
under the effect of microwave radiation in a GaAs quantum
well with AlAs/GaAs superlattice barriers is the same
as the nature of oscillations first observed in \cite{bib1} and
predicted in \cite{bib2}.

Note that, in our experimental dependences, unlike
\cite{bib6, bib29}, the point positioned near the cyclotron
resonance and characterized by the absence of the effect of
microwave radiation on the magnetoresistance of the
2D electron gas is shifted toward higher magnetic
fields. From Figs.~\ref{fig1} and \ref{fig2}, one can see that, at a
microwave frequency of $140$~$\mathrm{GHz}$, this shift is more
pronounced for the curves obtained at $4.2$~$\mathrm{K}$. We believe
that one of the possible reasons for such a shift is the
error in determining the magnitude of the magnetic
field component perpendicular to the surface of the 2D
electron gas; in our experiments, this error could reach
10\%. The problem of a precise position determination
for the specific points of magnetoresistance oscillations
induced by microwave radiation in 2D electron systems
with moderate mobility is still unsolved and will be the
subject of our subsequent experimental studies.

Today, most of the theoretical concepts explaining
the nature of these oscillations are based, on the one
hand, on indirect optical transitions accompanied by a
momentum variation due to the scattering by impurities
or phonons \cite{bib2,bib9,bib10,bib11,bib12} or, on the other hand,
on the nonequilibrium filling of electron states at the broadened
Landau levels \cite{bib7,bib14,bib15}. On the basis of experimental
data available to us by this day, we cannot decide
between the theoretical scenarios to explain the
oscillations arising in our samples under the effect of a
microwave field of the millimetric wave range. However, we
believe that one of the possible reasons for the
manifestation of the close-to-zero resistance state that is
induced by microwave radiation in GaAs quantum
wells with moderate mobility and high concentration of
2D electrons is the elastic electron scattering by rough
heteroboundaries with participation of photons. This
hypothesis is consistent with the role that is played by
the heteroboundaries of a GaAs quantum well with
AlAs/GaAs superlattice barriers in the manifestation of
magnetophonon resonance in this well \cite{bib30}.

Thus, we showed that, with an increase in the
concentration of the 2D electron gas, the zero-resistance
states induced by microwave radiation manifest themselves
in GaAs quantum wells with AlAs/GaAs superlattice
barriers for moderate mobility values. The
experimental data obtained by us make it possible to
shift the experimental studies of the nature of zero-resistance
states induced by electromagnetic field in 2D
electron systems to the submillimetric wave range and
to develop receivers for infrared radiation on the basis
of this effect.

We are grateful to I.V.~Marchishin for technical
assistance in the experiments. This work was supported
by the Russian Foundation for Basic Research (project
nos. 04-02-16789 and 06-02-16869) and by INTAS
(grant no. 03-51-6453).

\vfill\eject

\end{document}